\documentclass[11pt,a4paper,onecolumn,notitlepage]{article}

\usepackage{cmap} 

\usepackage[pdfpagemode=None,
pagebackref=true,
bookmarksopen=false,
hyperindex=true,
colorlinks=true,
linkcolor=blue,
citecolor=blue,
pdftitle={Categories of Br\`{e}gman operations and epistemic (co)monads},
pdfauthor={Ryszard Pawe{\l} Kostecki}]{hyperref}

\usepackage{tabularx} 
\usepackage{amsfonts} 
\usepackage{amsmath} 
\usepackage{mathrsfs} 
\usepackage{amssymb} 
\usepackage[OT2,T1]{fontenc} 
\usepackage[2cell,all]{xy} 

\DeclareMathAlphabet{\mathpzc}{OT1}{pzc}{m}{it}

\usepackage{geometry}
\usepackage{underscore}           

\usepackage{tabularx} 
\usepackage{amsfonts} 
\usepackage{amsmath} 
\usepackage{mathrsfs} 
\usepackage{amssymb} 
\usepackage[T1]{fontenc} 
\usepackage[2cell,all]{xy} 

\usepackage{setspace} 
\usepackage{titling}
\usepackage{changepage}

\usepackage{rpk}
\usepackage{bibspacing}
\usepackage{cyrillic}

\newcommand{\cyrrm}[1]{\mbox{\fontencoding{OT2}\fontfamily{wncyr}\selectfont#1}} 

\renewcommand*{\backref}[1]{}
\renewcommand*{\backrefalt}[4]{%
  \ifcase #1 %
    \relax
  \or
    $\uparrow$~#2.
  \else
    $\uparrow$~#2.
  \fi%
}

\begin{document}
\begin{center}
\begin{adjustwidth}{-3.8mm}{-3.8mm}
{\LARGE\textbf{Categories of Br\`{e}gman operations and epistemic (co)monads}}
\end{adjustwidth}
\vspace{4mm}{\Large Ryszard Pawe{\l} Kostecki}\\
\vspace{3mm}{\small \textit{National Quantum Information Center \& Institute of Informatics}}\\
\vspace{-1.2mm}{\small \textit{Faculty of Mathematics, Physics, and Informatics, University of Gda\'{n}sk}}\\
\vspace{-1.2mm}{\small \textit{Andersa 27, 81-824 Sopot, Poland}}\\
\vspace{0mm}{\small \textit{International Center for Theory of Quantum Technologies, University of Gda\'{n}sk}}\\
{\vspace{-1.2mm}\small \textit{Wita Stwosza 63, 80-308 Gda\'{n}sk, Poland}}\\
{\vspace{2mm}\small\texttt{kostecki@fuw.edu.pl}}\\
\vspace{3mm}{March 13, 2021}


\end{center}
\thispagestyle{empty}
%
%
\vspace{-1mm}
\begin{abstract}
\noindent We construct a categorical framework for nonlinear postquantum inference, with embeddings of convex closed sets of suitable reflexive Banach spaces as objects and pullbacks of Br\`{e}gman quasi-nonexpansive mappings (in particular, constrained maximisations of Br\`{e}gman relative entropies) as morphisms. It provides a nonlinear convex analytic analogue of Chencov's programme of geometric study of categories of linear positive maps between spaces of states, a working model of Mielnik's nonlinear transmitters, and a setting for nonlinear resource theories (with monoids of Br\`{e}gman quasi-nonexpansive maps as free operations, their asymptotic fixed point sets as free sets, and Br\`{e}gman relative entropies as resource monotones). We construct a range of concrete examples for semi-finite JBW-algebras and any W$^*$-algebras. Due to relative entropy's asymmetry, all constructions have left and right versions, with Legendre duality inducing categorical equivalence between their well-defined restrictions. Inner groupoids of these categories implement the notion of statistical equivalence. The hom-sets of a subcategory of morphisms given by entropic projections have the structure of partially ordered commutative monoids (so, they are resource theories in Fritz's sense). Further restriction of objects to affine sets turns Br\`{e}gman relative entropy into a functor. Finally, following Lawvere's adjointness paradigm for deductive logic, but with a semantic twist representing Jaynes' and Chencov's views on statistical inference, we introduce a category-theoretic multi-(co)agent setting for inductive inference theories, implemented by families of monads and comonads. We show that the br\`{e}gmanian approach provides some special cases of this setting.
\end{abstract}

\section{Introduction}\hypertarget{Section1}{}

This paper stems from the questions: 1) How to construct a theory of nonlinear (post)quantum operations, valid also in  continuously infinite dimensions, and exhibiting useful information-, resource-, and category-theoretic structures? 2) How to formalise the logics of inductive (predictive, statistical) inference theory, in (some) analogy to category-theretic formalisations of the logics of deductive inference?

Extension of the maximum entropy approach from model construction (objects) \cite{Elsasser:1937,Stratonovich:1955,Jaynes:1957,Jaynes:1957:2,Jaynes:1979:where:do:we:stand} to inductive inferences (morphisms) \cite{Sanov:1957,Kullback:1959,Bregman:1967,Chencov:1968,Hobson:1969} allows to derive the Bayes--Laplace and Jeffrey's rules \cite{Williams:1980,Warmuth:2005,Caticha:Giffin:2006,Douven:Romeijn:2012}, L\"{u}ders' and quantum Jeffrey's rules \cite{Hellmann:Kaminski:Kostecki:2016}, partial trace \cite{MunkNielsen:2015}, and (preduals of) conditional expectations \cite{Kostecki:Munk:2021} as special cases of constrained maximisation of the Kullback--Leibler/Umegaki relative entropy (left or right $D_1$-projections). However, to establish a full-fledged relative entropic (post)bayesian setting, two related problems have to be solved: 1) choice of a sufficiently rich and well-behaved class of relative entropies; 2) identification of a family of morphisms, which (roughly) could be for entropic projections what CPTP maps are for (preduals of) conditional expectations. Br\`{e}gman \cite{Bregman:1967} introduced a class $-D_\Psi$ of relative entropies satisfying a generalisation of a pythagorean equation with respect to left $D_\Psi$-projections (independently, Chencov \cite{Chencov:1968} discovered its right version for $D_1$). The works \cite{Alber:1993,Alber:Butnariu:1997,Butnariu:Iusem:1997,Butnariu:Iusem:2000,Bauschke:Borwein:Combettes:2001,Bauschke:Borwein:Combettes:2003,Resmerita:2004} (among others) established a successful Banach space generalisation of Br\`{e}gman's theory. Reich \cite{Reich:1996} introduced a class of left strongly $D_\Psi$-quasi-nonexpansive maps, LSQ$(\Psi)$, which is closed under composition, and (under some additional conditions \cite{Bauschke:Borwein:Combettes:2003,Reich:Sabach:2011}) contains left $D_\Psi$-projections. Right versions of these results also hold \cite{MartinMarquez:Reich:Sabach:2012,MartinMarquez:Reich:Sabach:2013:BSN}. Right and left $D_\Psi$-projections are generalisations of Hilbert space metric projections, better behaved and weaker than Banach space metric projections \cite{Alber:1993,Alber:1996}. They are characterised by the corresponding pythagorean equations, while the criteria of convergence defining LSQ$(\Psi)$ and RSQ$(\Psi)$ are, in essence, topological versions of these equations. However, good behaviour of these maps requires some additional geometric properties of the underlying Banach space $X$. These properties do not hold neither for generic base norm spaces nor for preduals of JBW- and W$^*$-algebras. The missing link, provided by us in \cite{Kostecki:2017}, is twofold: 1) introducing Br\`{e}gman $\ell$-information $D_{\ell,\Psi}$ by a bijective pullback ($\ell$-embedding) of $D_\Psi$ from geometrically well behaved (e.g., reflexive) space $X$, constructed over not so well behaved (e.g., base norm) spaces (and doing the same with $\cdot$SQ$(\Psi)$ maps); 2) providing rich family of models (i.e., triples $(X,\ell,\Psi)$), by: 2a) establishing characterisation of legendreness, and a sufficient condition for LSQ-adaptedness and RSQ-composability, of a family $\Psi_\varphi:=\int_0^{\n{\cdot}_X}\dd t\varphi(t)$, where $\varphi$ is a gauge function of a nonlinear duality map $\DG\Psi_\varphi=j_\varphi:X\ra X^\star$ \cite{Beurling:Livingston:1962,Asplund:1967}; 2b) constructing a range of concrete models in nonassociative and noncommutative settings.\footnote{In \hyperlink{Section2}{\S2} we recall basic notions of convex nonlinear analysis and br\`{e}gmanian theory in reflexive Banach spaces, discussed in details, with further references, in \cite{Borwein:Vanderwerff:2010,Sabach:2012,Kostecki:2017}. As for \hyperlink{Section3}{\S3}, the notions of Br\`{e}gman $\ell$-information and $\ell$-projection were introduced (abstractly) in \cite{Kostecki:2011:OSID} (with $\ell$-embeddings generalising earlier ideas of \cite{Mazur:1929,Kaczmarz:1933} and \cite{Nagaoka:Amari:1982,Kosaki:1984:uniform,Zhu:Rohwer:1997,Gibilisco:Pistone:1998,Gibilisco:Isola:1999,Raynaud:2002,Jencova:2005}), and are studied (concretely, with the corresponding $D_\Psi$-quasi-nonexpansive $\ell$-operations, including the examples of \hyperlink{Section4}{\S4}, as well as an extension to nonreflexive case, left and right $D_\Psi$-Chebysh\"{e}v sets, continuity of $D_\Psi$-projections, limitations of Legendre duality, etc) in \cite{Kostecki:2017}. The rest of this paper is new, and can be seen as a category-theoretic counterpart to \cite{Kostecki:2017}.} The resulting families LSQ$(\ell,\Psi)$ and RSQ$(\ell,\Psi)$ provide nonlinear convex analytic analogues of linear CPTP maps, based on the geometry of generalised pythagorean equation, as opposed to tensor products. While $\ell$-embeddings of br\`{e}gmanian structures solve a mathematical problem, they introduce a conceptual one: the basic objects (information state spaces) of a theory are $\ell$-closed $\ell$-convex sets, which, as opposed to compact convex sets, do not rely on base norm/order unit semantics (allowing for ``information theory without probability'' (c.f. \cite{Ingarden:Urbanik:1962}) on objects and hom-sets of categories which do not admit any (generalised) probabilistic structure).\footnote{Our focus on categories of inductive inference morphisms, instead of axiomatisation of probability spaces, follows the insights: \cytat{Many physicists take it for granted that their theories can be either refuted or verified by comparison with experimental data. In order to evaluate such data, however, one must employ statistical estimation and inference methods which, unfortunately, always involve an ad hoc proposition. (...) no verification is possible unless the relevant inference method is an integral part of the theory} \cite{Sykora:1974} (c.f. also \cite{Giles:1970}), \cytat{the motion creates the form} \cite{Mielnik:1981:motion} (c.f. also \cite{Davies:1974:symmetries}), and \cytat{[t]he main goal of statistician is to choose a priori reasonable families guaranteeing good rates of convergence of loss functions} \cite{Chencov:1987}. In addition, we see the passage from `linearity' (in the sense of \cite{Mielnik:1969,Davies:Lewis:1970,Gudder:1973,Mielnik:1974}) to `nonlinearity' (in our sense) along the lines of: \cytat{the great watershed in optimization isn't between linearity and nonlinearity, but convexity and nonconvexity} \cite{Rockafellar:1993} (c.f. also \cite{Lawvere:2000}).} This relativisation of the type of optimal/ideal experimental data with respect to the choice of the system of inductive inference (c.f. \cite{Domotor:1985}) requires to be coherently addressed. For this purpose, in \hyperlink{Section8}{\S8} we propose a categorical approach to adjointness between theoretical model construction and predictive verification, modeled after Lawvere's approach to categorical deductive logic \cite{Lawvere:1963}, yet with a twist, taking into account Chencov's and Jaynes' approaches to mathematical and conceptual foundations of statistical inference. We show that, under some conditions, it forms a resource theory of intersubjective knowledge (with agency of free operations and coagency of selection of referential experimental designs).

\noindent\section{Br\`{e}gman projections and quasi-nonexpansive maps}\hypertarget{Section2}{}

In terminological (resp., mathematical) agreement with \cite{Wiener:1948,Renyi:1961,Umegaki:1961,Bratteli:Robinson:1979} (resp., \cite{Eguchi:1983,Csiszar:1995}), we define: an \df{information} on a set $Z$ as a map $D:Z\times Z\ra[0,\infty]$ such that $D(x,y)=0$ $\iff$ $x=y$; a \df{relative entropy} as $-D$. Given a function $f:Y\ra]-\infty,\infty]$ on a real Banach space $Y$ with $\efd(f):=\{x\in Y\mid f(x)\neq\infty\}\neq\varnothing$, $f^\lfdual$ will denote a Fenchel dual of $f$ with respect to a bilinear duality map $\duality{x,y}:=y(x)\in\RR$ $\forall(x,y)\in Y\times Y^\star$, where $Y^\star$ denotes the Banach dual of $Y$. In what follows, $X$ denotes a reflexive real Banach space, $\INT$ denotes an interior in norm topology of $X$, and $\Psi:X\ra]-\infty,\infty]$ is Legendre \cite{Rockafellar:1967,Bauschke:Borwein:Combettes:2001} (so, its Gateaux derivative is a bijection, $\DG\Psi:\intefd{\Psi}\ra\intefd{\Psi^\lfdual}$, with $(\DG\Psi)^{-1}=\DG\Psi^\lfdual$). A map $D_\Psi:X\times X\ra[0,\infty]$, $D_\Psi(z,w):=\Psi(z)-\Psi(w)-\duality{z-w,\DG\Psi(w)}=\Psi(z)+\Psi^\lfdual(\DG\Psi(w))-\duality{z,\DG\Psi(w)}$ for $w\in\intefd{\Psi}$ and $\infty$ otherwise \cite{Brunk:Ewing:Utz:1957,Bregman:1967,Kiwiel:1997:proximal,Butnariu:Iusem:1997}, is an information \cite{Bauschke:Borwein:Combettes:2001}, called \df{Br\`{e}gman information}. For $y\in\intefd{\Psi}$, $C\subseteq X$, and $\varnothing\neq C\cap\intefd{\Psi}$, if the set $\arginff{x\in C}{D_\Psi(x,y)}$ (resp., $\arginff{x\in C}{D_\Psi(y,x)}$) is a singleton, then its element will be denoted $\LPPP^{D_\Psi}_C(y)$ (resp., $\RPPP^{D_\Psi}_C(y)$), and called \df{left} (resp., \df{right}) \df{$D_\Psi$-projection} of $y$ onto $C$. Both left and right $D_\Psi$-projections are idempotent. If $\varnothing\neq C\subseteq\intefd{\Psi}$ is convex and closed, then $\forall y\in\intefd{\Psi}$ $\exists!\LPPP^{D_\Psi}_C(y)$ \cite{Bauschke:Borwein:Combettes:2001}. Furthermore, $\RPPP^{D_\Psi}_K=\DG\Psi^\lfdual\circ\LPPP^{D_{\Psi^\lfdual}}_{\DG\Psi(K)}\circ\DG\Psi$, and $\RPPP^{D_{\Psi^\lfdual}}_{\DG\Psi(M)}=\DG\Psi\circ\LPPP^{D_\Psi}_M\circ\DG\Psi^\lfdual$ for nonempty, closed, convex sets $\DG\Psi(K)$ and $M$ \cite{Bauschke:Wang:Ye:Yuan:2009,Bauschke:Macklem:Wang:2011,MartinMarquez:Reich:Sabach:2012,Luo:Meng:Wen:Yao:2019}. If $K$ is a closed affine subspace of $X$, then the \df{left pythagorean equation}, $D_\Psi(x,y)=D_\Psi(x,\LPPP^{D_\Psi}_K(y))+D_\Psi(\LPPP^{D_\Psi}_K(y),y)$ $\forall(x,y)\in K\times\intefd{\Psi}$, holds \cite{Bregman:1967,Alber:Butnariu:1997,Alber:2007}. If $\DG\Psi(K)$ is a closed affine subspace of $X$, then the \df{right pythagorean equation}, $D_\Psi(x,y)=D_\Psi(x,\RPPP^{D_\Psi}_K(x))+D_\Psi(\RPPP^{D_\Psi}_K(x),y)$ $\forall (x,y)\in\intefd{\Psi}\times K$, holds \cite{Chencov:1968,MartinMarquez:Reich:Sabach:2012}. If $K$ is convex instead of affine, then `$=$' in these two equations turns into `$\geq$'.

Given $\varnothing\neq M\subseteq\intefd{\Psi}$ and a function $T:M\ra\intefd{\Psi}$, $\Fix(T):=\{x\in M\mid T(x)=x\}\neq\varnothing$ is called a set of \df{fixed points}, while $\widehat{\Fix}(T)$, called a set of \df{asymptotic fixed points} consists of such $x\in M$ that there exists a sequence $\{x_n\}_{n\in\NN}\subseteq M$ weakly convergent to $x$ with $\lim_{n\ra\infty}\n{x_n-Tx_n}_X=0$. In general, $\Fix(T)\subseteq\widehat{\Fix}(T)$. $T:M\ra\intefd{\Psi}$ is called: \df{completely $D_\Psi$-nonexpansive} (or \df{CN$(\Psi)$}) \cite{Butnariu:Iusem:2000} if{}f $D_\Psi(T(x),T(y))\leq D_\Psi(x,y)$ $\forall x,y\in M$; \df{left strongly $D_\Psi$-quasi-nonexpansive} (or \df{LSQ$(\Psi)$}) \cite{Censor:Reich:1996,Reich:1996,MartinMarquez:Reich:Sabach:2013:BSN} if{}f $D_\Psi(x,T(y))\leq D_\Psi(x,y)$ $\forall(x,y)\in\widehat{\Fix}(T)\times M$ and $(p\in\widehat{\Fix}(T)$, $\{y_n\}_{n\in\NN}$ is bounded, $\lim_{n\ra\infty}(D_\Psi(p,y_n)$ $-D_\Psi(p,Ty_n))=0)$ $\limp$ $\lim_{n\ra\infty}D_\Psi(Ty_n,y_n)=0$; \df{right strongly $D_\Psi$-quasi-nonexpansive} (or \df{RSQ$(\Psi)$}) \cite{MartinMarquez:Reich:Sabach:2012} if{}f $D_\Psi(T(x),y)\leq D_\Psi(x,y)$ $\forall(x,y)\in M\times\widehat{\Fix}(T)$ and ($p\in\widehat{\Fix}(T)$, $\{y_n\}_{n\in\NN}$ is bounded, $\lim_{n\ra\infty}$ $(D_\Psi(y_n,p)-D_\Psi(T(y_n),p))=0$) $\limp$ $\lim_{n\ra\infty}(y_n,T(y_n))=0$. In general, $\LPPP^{D_\Psi}_C$ is not CN$(\Psi)$. If $\Psi$ is bounded, supercoercive, (uniformly Fr\'{e}chet differentiable and totally convex) on bounded subsets of $X$,  $\varnothing\neq K\subseteq\intefd{\Psi}$, $\{T_1,\ldots,T_n\}$ are LSQ$(\Psi)$ functions $K\ra K$ such that $\widehat{F}:=\bigcap_{i=1}^n\widehat{\Fix}(T_i)\neq\varnothing$ and $T:=T_n\circ\cdots\circ T_1$, then $\widehat{\Fix}(T)\subseteq\widehat{F}$, and if $\widehat{\Fix}(T)\neq\varnothing$ then $T$ is LSQ$(\Psi)$ \cite{Reich:1996,Sabach:2012,MartinMarquez:Reich:Sabach:2013:BSN}. Such $\Psi$ will be called \df{LSQ-compositional}. If, additionally, $\efd(\Psi)=X$, then we will call it \df{LSQ-adapted}. If $\Psi:X\ra\RR$ is (bounded, uniformly continuous, and totally convex) on bounded subsets of $X$, $\varnothing\neq K\subseteq X$, $\{T_1,\ldots,T_n\}$ are RSQ$(\Psi)$ functions $K\ra K$ such that $\widehat{F}:=\bigcap_{i=1}^n\widehat{\Fix}(T_i)\neq\varnothing$ and $T:=T_n\circ\cdots\circ T_1$, then $\widehat{\Fix}(T)\subseteq\widehat{F}$, and if $\widehat{\Fix}(T)\neq\varnothing$ then $T$ is RSQ$(\Psi)$ \cite{MartinMarquez:Reich:Sabach:2013:BSN}. Such $\Psi$ will be called \df{RSQ-compositional}. If, additionally, $\Psi$ is totally convex on $X$, $\Psi^\lfdual$ is totally convex on $\intefd{\Psi^\lfdual}$, and $\DG\Psi$ is weakly sequentially continuous, then we will call $\Psi$ \df{RSQ-adapted}. The results of \cite{Bauschke:Borwein:Combettes:2003,Reich:Sabach:2011,MartinMarquez:Reich:Sabach:2012} imply \cite{Kostecki:2017}: (i) For any LSQ-adapted $\Psi$ and nonempty closed convex $K\subseteq\intefd{\Psi}$, $\widehat{\Fix}(\LPPP^{D_\Psi}_K)=\Fix(\LPPP^{D_\Psi}_K)=K$, hence $\LPPP^{D_\Psi}_K$ is LSQ$(\Psi)$; (ii) For any RSQ-adapted $\Psi$ and closed convex $\varnothing\neq M\subseteq\intefd{\Psi}$, $\widehat{\Fix}(\RPPP^{D_\Psi}_{\DG\Psi^\lfdual(M)})=\Fix(\RPPP^{D_\Psi}_{\DG\Psi^\lfdual(M)})=\DG\Psi^\lfdual(M)$, hence $\RPPP^{D_\Psi}_{\DG\Psi^\lfdual(M)}$ is RSQ$(\Psi)$.

\noindent\section{$\ell$-operations and nonlinear resource theories of states}\hypertarget{Section3}{}

Given a set $Z\subseteq\intefd{\Psi}$, a set $U$, and a bijection $\ell:U\ra Z$, we define the \df{Br\`{e}gman $\ell$-information} on $U$ as $D_{\ell,\Psi}(\phi,\psi):=D_{\Psi}(\ell(\phi),\ell(\psi))$ $\forall \phi,\psi\in U$. The properties of $D_\Psi$ can be naturally extended to the properties of $D_{\ell,\Psi}$, by turning $\ell$ into a homeomorphism. Given $C\subseteq U$, if $\ell(C)$ is convex (resp., affine; closed), then $C$ will be called \df{$\ell$-convex} (resp., \df{$\ell$-affine}; \df{$\ell$-closed}). So, the $\ell$-closure of $C\subseteq U$ is a closure of $C$ in the topology induced by $\ell$ from the norm topology of $X$. A \df{left} (resp., \df{right}) \df{$D_{\ell,\Psi}$-projection} is defined by $\LPPP^{D_{\ell,\Psi}}_C(\psi):=\LPPP^{D_\Psi}_{\ell(C)}(\ell(\psi))$ (resp., $\RPPP^{D_{\ell,\Psi}}_C(\psi):=\RPPP^{D_\Psi}_{\ell(C)}(\ell(\psi))$ for any $\ell$-closed $\ell$-convex (resp., $(\DG\Psi\circ\ell)$-closed $(\DG\Psi\circ\ell)$-convex) set $C$ and any $\psi\in U$. For $\varnothing\neq W\subseteq U$ and $T:\ell(W)\ra Z$, $T^\ell:\ell^{-1}\circ T\circ\ell:W\ra U$ will be called an \df{$\ell$-operation} (or an \df{$\ell$-transmitter}). The classes of $D_\Psi$-quasi-nonexpansive maps on $X$ determine the corresponding classes of $\ell$-operations on $U$ (i.e., \df{$(\ell,\Psi)$-transmitters}), in particular: CN$(\ell,\Psi)$, LSQ$(\ell,\Psi)$, and RSQ$(\ell,\Psi)$. We will denote $\widehat{\Fix}(T^\ell):=\ell^{-1}(\widehat{\Fix}(T))$. Each $y\in\intefd{\Psi^\lfdual}$ defines an \df{$(\ell,\Psi)$-observable} on $U$, given by $y\circ\ell:U\ra\RR$.\footnote{If $W$ is a convex set, then an $\ell$-operation is belongs to Mielnik's nonlinear transmitters \cite{Mielnik:1969,Mielnik:1974}. In our case, $\ell$-convexity of $W$ is more fundamental property then convexity, so $T$ plays the role of a nonlinear transmitter, with $(X,X^\star)$ \textit{linearly} representing (states/sources/resources, observables/sinks/witnesses), while $T^\ell$ is its nonlinear representation. The monoids of $\ell$-operations can be seen as analogues of Mielnik's semigroups of mobility \cite{Mielnik:1981:motion}, while $(\ell,\cdot)$-observables provide an instance of Mielnik's observables \cite{Mielnik:1974}. So, (ii$^{\mathrm{L},\mathrm{R}}_{\ell,\Psi}$) provide a weakened version of Mielnik's nonlinear generalised quantum theory. (Mielnik identifies observables with \textit{any} maps $f:U\ra\RR$, if $U$ is convex. In our case one may consider such $f$ as an `intersubjective' observable, which is sampled in terms of `subjective' $(\ell,\Psi)$-observables; the same goes for nonlinear transmitters as well as $U$ itself. \hyperlink{Section8Ex2}{\S8.Ex.2} provides further development of this dialectics.)}

Given a set $U$ (of states), we define a \df{resource theory of states} (c.f., e.g., \cite{Horodecki:Horodecki:Horodecki:Horodecki:2009,delRio:Kraemer:Renner:2015,Chitambar:Gour:2019,Takagi:Regula:2019}) as a triple $(P,S,R)$, where $P$ is a submonoid of a monoid $\mathrm{End}(U)$ of endomorphisms of $U$, $\varnothing\neq S\subseteq U$ satisfies $P(S)\subseteq S$, and $R:=\{r:U\ra\RR^+\mid(r\circ p)(\phi)\leq r(\phi)\;\forall\phi\in U\;\forall p\in P\}$. The elements of $P$ (resp., $S$; $R$) are called \df{free operations} (resp., \df{free states}; \df{resource monotones}). For example, in our setting, we have:
\begin{enumerate}
\item[(i$^{\mathrm{L}/\mathrm{R}}_{\ell,\Psi}$)] $(\mathcal{T},S_{\mathcal{T}},\{D^{\mathrm{L}}_{S_{\mathcal{T}}}\})$ (resp., $(\mathcal{T},S_{\mathcal{T}},\{D^{\mathrm{R}}_{S_{\mathcal{T}}}\})$): if $\mathcal{T}$ is a submonoid of $\mathrm{CN}(\ell,\Psi)$ with $\mathcal{T}(S_{\mathcal{T}})\subseteq S_{\mathcal{T}}$, and $\varnothing\neq S_{\mathcal{T}}\subseteq U$ is $\ell$-closed $\ell$-convex (resp., $(\DG\Psi\circ\ell)$-closed $(\DG\Psi\circ\ell)$-convex), then $D^{\mathrm{L}}_{S_{\mathcal{T}}}:=\inf_{\phi\in S_{\mathcal{T}}}\{D_{\ell,\Psi}(\phi,\,\cdot\,)\}$ (resp., $D^{\mathrm{R}}_{S_{\mathcal{T}}}:=\inf_{\phi\in Q_{\mathcal{T}}}\{D_{\ell,\Psi}(\,\cdot\,,\phi)\}$) is a resource monotone; an interesting example is given by $S_{\mathcal{T}}=\{\phi\in U\mid\forall\psi\in U\exists T\in\mathcal{T}\;\;T(\psi)=\phi\}\neq\varnothing$;
\item[(ii$^{\mathrm{L}/\mathrm{R}}_{\ell,\Psi}$)] $(\mathcal{T},\widehat{\Fix}(\mathcal{T}),\bigcup_{\phi\in\widehat{\Fix}(\mathcal{T})}\{D_{\ell,\Psi}(\phi,\,\cdot\,)\})$ (resp., $(\mathcal{T},\widehat{\Fix}(\mathcal{T}),\bigcup_{\phi\in\widehat{\Fix}(\mathcal{T})}\{D_{\ell,\Psi}(\,\cdot\,,\phi)\})$): if $\Psi$ is LSQ-(resp., RSQ-)compositional, $\varnothing\neq K\subseteq U$, $\mathcal{T}\subseteq\mathrm{LSQ}(\ell,\Psi)$ (resp., $\mathcal{T}\subseteq\mathrm{RSQ}(\ell,\Psi)$) is a monoid such that $T^\ell:K\ra K$ $\forall T^\ell\in\mathcal{T}$, $\bigcap_{i=1}^n\widehat{\Fix}(T_i)\neq\varnothing$ and $\widehat{\Fix}(T_1\circ\cdots\circ T_n)\neq\varnothing$ $\forall\{T^\ell_1,\ldots,T^\ell_n\}\subseteq\mathcal{T}$, then $D_{\ell,\Psi}(\phi,\,\cdot\,)$ (resp., $D_{\ell,\Psi}(\,\cdot\,,\phi)$) is a resource monotone for any $\phi\in\widehat{\Fix}(\mathcal{T})$; 
\item[(iii$^{\mathrm{L}/\mathrm{R}}_{\ell,\Psi}$)] $(\mathcal{T},K,\bigcup_{\phi\in K}\{D_{\ell,\Psi}(\phi,\,\cdot\,)\})$ (resp., $(\mathcal{T},K,\bigcup_{\phi\in K}\{D_{\ell,\Psi}(\,\cdot\,,\phi)\})$): for any $\ell$-closed $\ell$-convex (resp., $(\DG\Psi\circ\ell)$-closed $(\DG\circ\ell)$-convex) set $\varnothing\neq K\subseteq U$, if $\mathcal{T}$ is given by the set of all $\LPPP^{D_{\ell,\Psi}}_Q$ (resp., $\RPPP^{D_{\ell,\Psi}}_Q$) with $\ell$-closed $\ell$-convex (resp., $(\DG\Psi\circ\ell)$-closed $(\DG\Psi\circ\ell)$-convex) $Q$ such that $K\subseteq Q$, equipped with the composition $\LPPP^{D_{\ell,\Psi}}_{Q_1}\diamond\LPPP^{D_{\ell,\Psi}}_{Q_2}:=\LPPP^{D_{\ell,\Psi}}_{Q_1\cap Q_2}$ (resp., $\RPPP^{D_{\ell,\Psi}}_{Q_1}\diamond\RPPP^{D_{\ell,\Psi}}_{Q_2}:=\RPPP^{D_{\ell,\Psi}}_{Q_1\cap Q_2}$).
\end{enumerate}
In these examples $D_{\ell,\Psi}$ plays three different roles: it provides resource monotones, controls the behaviour of free operations, and participates in the construction of a space of free states. If $\Psi$ is LSQ-adapted (resp., RSQ-adapted), then (iii$^{\mathrm{L}}_{\ell,\Psi}$) (resp., (iii$^{\mathrm{R}}_{\ell,\Psi}$)) is a special case of (ii$^{\mathrm{L}}_{\ell,\Psi}$) (resp., (ii$^{\mathrm{R}}_{\ell,\Psi}$)). Limited structural benefits of $\mathrm{CN}(\Psi)$ maps\footnote{\cytat{This generalization does not satisfy any of the properties that the classical nonexpansive operators do} \cite{Sabach:2012}. However, see \hyperlink{Section4Ex2b}{\S4.Ex.2(b)} and its consequence in \hyperlink{Section5}{\S5}, providing a nontrivial intersection of $(\ell,\Psi)$-transmitters $T^\ell$ with CPTP maps (on preduals of any W$^*$-algebras $\N$). This shall be considered in parallel to a characterisation of (preduals of) conditional expectations on (finite dimensional) $\N$ as right $D_1$-projections \cite{Kostecki:Munk:2021}, which, combined with \hyperlink{Section4Ex4}{\S4.Ex.4}, turns submonoids of (preduals of) conditional expectations into special cases of the type (iii$^{\mathrm{R}}_{\ell,\Psi}$) resource theory. Taking into account Chencov's geometric approach \cite{Chencov:1972} to Wald's statistical decision theory \cite{Wald:1939,Wald:1950} (see \hyperlink{Section8}{\S8}), and Holevo's approach \cite{Holevo:1972:analog,Holevo:1973:aspekty} to selection of POVM as a minimiser of a quantum statistical decision rule, one can view $T^\ell$ can as an analogue of a statistical decision rule.} 
 allow to consider: $\widehat{\Fix}(\mathcal{T})$ as a generic notion of a free state space in br\`{e}gmanian resource theories; elements of $\mathrm{LSQ}(\ell,\Psi)\setminus\mathcal{T}$ and $\mathrm{RSQ}(\ell,\Psi)\setminus\mathcal{T}$ as the generic nonfree operations. The (linear) \df{witnesses} of $S$ are defines as the elements of $\{y\in\intefd{\Psi^\lfdual}\mid\duality{x,y}\geq0\,\forall x\in S\}$. Using \cite{MartinMarquez:Reich:Sabach:2013:BSN}, we observe that any set $\{\mathcal{T}_1,\ldots,\mathcal{T}_m\}$, $m\in\NN$, of resource theories of the type (ii$^{\mathrm{R}}_{\ell,\Psi}$) admits a convex envelope $\co(\mathcal{T}_1,\ldots,\mathcal{T}_m):=\{T^\ell\in\mathrm{RSQ}(\ell,\Psi)\mid T:=\sum_{i=1}^nw_iT_i,\;\sum_{i=1}^nw_i=1,\;\{T^\ell_1,\ldots,T^\ell_n\}\in\mathcal{T}_1\cup\cdots\cup\mathcal{T}_m,\;(w_1,\ldots,w_n)\in]0,1[^n\}$, satisfying $D_{\ell,\Psi}(T^\ell(\psi),\phi)\leq\sum_{i=1}^nw_i D_{\ell,\Psi}(T^\ell_i(\psi),\phi)$ $\forall\psi,\phi\in K$. So, while $\co(\mathcal{T}_1,\ldots,\mathcal{T}_m)$ may be not a monoid (hence, not a resource theory of states), it provides a setting of a \df{multi-resource theory of states}, with its elements understood as (generally, nonfree) operations, decomposable (``tomographable'') into weighted mixtures of free operations from different individual resource theories. See \cite{Buscemi:Kostecki:2021} for further br\`{e}gmanian resource theoretic discussion.

\noindent\section{Examples of $D_{\ell,\Psi}$}\hypertarget{Section4}{}

\textbf{Ex.1.} \textbf{(a)} For any Banach space $X$, a \df{duality map} is defined as $j_\varphi:X\ni x\mapsto\{y\in X^\star\mid\duality{x,y}=\n{x}_X\n{y}_{X^\star}, \n{y}_{X^\star}=\varphi(\n{x}_X)\}\subseteq X$ \cite{Beurling:Livingston:1962}, where $\varphi:\RR^+\ra\RR^+$, called a \df{gauge}, is strictly increasing, continuous, with $\varphi(0)=0$ and $\lim_{t\ra\infty}\varphi(t)=\infty$. For any gauge $\varphi$, $\Psi_\varphi:=\int_0^{\n{\cdot}_X}\dd t\varphi(t):X\ra\RR^+$ is continuous, convex, and increasing \cite{Asplund:1967,Zalinescu:1983}. If $X$ is Gateaux differentiable, then $j_\varphi=\{\DG\Psi_\varphi\}$ \cite{Asplund:1967}. E.g., $\Psi_\varphi=\beta\n{\cdot}_X^{1/\beta}$ (resp., $\n{\cdot}_X$) for $\varphi(t)=t^{\frac{1}{\beta}-1}$ (resp., $\frac{1}{\beta}t^{\frac{1}{\beta}-1}$) with $\beta\in\,]0,1[$. In \cite{Kostecki:2017}, using and extending the results of \cite{Zalinescu:1983,Butnariu:Iusem:2000,Bauschke:Borwein:Combettes:2001,Zalinescu:2002,Resmerita:2004}, we prove: 1) $\Psi_\varphi$ is Legendre if{}f $X$ is Gateaux differentiable and strictly convex; 2) $\Psi_\varphi$ is LSQ-adapted and RSQ-compositional (and $\LPPP^{D_{\Psi_\varphi}}$ is norm-to-norm continuous \cite{Resmerita:2004}) if $X$ is locally uniformly convex and uniformly Fr\'{e}chet differentiable.\footnote{RSQ-adaptedness of $\Psi_\varphi$ requires weak sequential continuity of $j_\varphi=\DG\Psi_\varphi$, which is an underinvestigated property, yet known to hold for $\varphi(t)=t^{1/\beta-1}$ on $l_{1/\beta}$ sequence spaces with $\beta\in\,]0,1[$ \cite{Browder:1966} and on Hilbert spaces for $\beta=\frac{1}{2}$ \cite{Xu:Kim:Yin:2014}.} \textbf{(b)} For any base norm space $Y$, it is reflexive if{}f its base is weakly compact (see, e.g., \cite{Alfsen:Shultz:2003}). In such case, some results of br\`{e}gmanian theory apply directly, under weakening of $\DG\Psi$ to right Gateaux derivative (so, Legendre duality is lost), and with $\ell=\id_Y$ (more generally, $\ell$ be taken to be any automorphism of $Y$). This holds, in particular, for any finite dimensional $Y$, for type I$_2$ JBW-factors \cite{Topping:1965} (which are exactly the spin factors $\H\oplus\RR$, where $\H$ is a Hilbert space with $\dim\H\geq2$ \cite{Stoermer:1966:Jordan}), as well as for state spaces of orthomodular posets satisfying Jordan--Hahn decomposition property \cite{Fischer:Ruettiman:1978:geometry}. \textbf{(c)} Given any base norm space $Y$, if $U$ is a generating positive cone of $Y$ and $\ell$ is a map from $U$ (or $Y$) into a geometrically well-behaved Banach space $X$, then $D_\Psi$ determines $D_{\ell,\Psi}$ on $U$ (or $Y$) and thus also on the base of $Y$. Ex.2 and Ex.3 provide the special cases of this situation, with $X$ implementing Ex.1(a) (for any W$^*$-algebra $\N$, using the uniquness of a polar decomposition, we extend the bijective embedding to the whole predual space $\N_\star$,  under replacement of $\duality{\cdot,\cdot}$ by $\re\duality{\cdot,\cdot}$ in all formulas), calling for further investigation of convexity and differentiability of Banach spaces of integrals over general base norm spaces \cite{Tikhonov:1990,Tikhonov:1992,Tikhonov:1993}. 

\textbf{Ex.2.} \textbf{\hypertarget{Section4Ex2a}{(a)}} If $A$ is a semi-finite JBW-algebra with a faithful normal semi-finite trace $\tau:A^+\ra[0,\infty]$, then the nonassociative $L_{1/\gamma}$ spaces, $\gamma\in\,]0,1]$, defined by $(L_{1/\gamma}(A,\tau):=\overline{A_\tau}^{\n{\cdot}_{1/\gamma}},\n{x}_{1/\gamma}:=(\tau(\ab{x}^{1/\gamma}))^\gamma)$, where $A_\tau:=\Span\{x\in A^+\mid\tau(x)<\infty\}$ \cite{Abdullaev:1983,Abdullaev:1984,Iochum:1984}, are uniformly convex and uniformly Fr\`{e}chet differentiable for $\gamma\in\,]0,1[$ \cite{Iochum:1986}. Hence, for any gauge $\varphi$, $\Psi_{\varphi,\gamma}:=\int_0^{\n{\cdot}_{1/\gamma}}\dd t\varphi(t)$ is LSQ-adapted and RSQ-compositional. By means of $\phi=\tau(h_\phi\circ\cdot)$, the nonassociative Mazur map $\ell_\gamma:A_\star\ni\phi\mapsto \sign(h_\phi)\ab{h_\phi}^\gamma\in L_{1/\gamma}(A,\tau)$ determines  $D_{\gamma,\varphi}:=D_{\ell_\gamma,\Psi_{\varphi,\gamma}}:A_\star\times A_\star\ra[0,\infty]$. Due to isometric isomorphism of $L_{1/\gamma}(A,\tau)$ for different $\tau$ \cite{Ayupov:Abdullaev:1989}, $D_{\gamma,\varphi}$ do not depend on $\tau$. For $(\beta,\gamma)\in\,]0,1[^2$ and $\omega,\phi\in\A_\star^+$, $\varphi(t)=t^{1/\beta-1}/\beta$ yields $D_{\gamma,\beta}(\omega,\phi)=(\tau(h_\omega))^{\gamma/\beta}+\frac{1}{1-\beta}(\tau(h_\phi))^{\gamma/\beta}-\frac{1}{\beta}(\tau(h_\phi))^{\gamma/\beta-1}\tau(h_\omega^\gamma\circ h_\phi^{1-\gamma})$ for $\omega\ll\phi$ and $\infty$ otherwise. \textbf{\hypertarget{Section4Ex2b}{(b)}} The same (including the formula for $D_{\gamma,\beta}$) holds for any W$^*$-algebra $\N$ and $\gamma\in]0,1[$, due to uniform convexity and uniform Fr\'{e}chet differentiability of $(L_{1/\gamma}(\N),\n{\cdot}_{1/\gamma})$ spaces \cite{Terp:1981,Masuda:1983,Kosaki:1984:ncLp}, under replacement of $\phi=\tau(h_\phi\circ\cdot)$ by $\phi=\int\phi\,\cdot\,$, and with $\ell_\gamma:\N_\star\ni\phi\mapsto u_\phi\ab{\phi}^\gamma\in L_{1/\gamma}(\N)$, where $\phi=\ab{\phi}(\cdot u_\phi)$ is a polar decomposition, while $\int$ and $\ab{\phi}^\gamma$ are defined as in \cite{Falcone:Takesaki:2001}. For $\varphi(t)=t^{(1-\gamma)/\gamma}/(\gamma(1-\gamma))=:\varphi_\gamma(t)$ we obtain $D_\gamma(\omega,\phi)=(\gamma(1-\gamma))^{-1}\int(\gamma\omega+(1-\gamma)\phi-\omega^\gamma\phi^{1/\gamma})$ whenever $\omega\ll\phi$ and $\infty$ otherwise, introduced in \cite{Jencova:2005,Kostecki:2011:OSID}, and unifying $D_\gamma$ of \cite{Liese:Vajda:1987,Hasegawa:1993}. All CPTP maps are completely $D_\gamma$-nonexpansive \cite{Jencova:2005}, so, the resource theories (i$^{\mathrm{L},\mathrm{R}}_{\ell_\gamma,\Psi_{\varphi_\gamma}}$) are valid submonoids (hence, resource theories) of CPTP maps.

\textbf{\hypertarget{Section4Ex3}{Ex.3}.} Given a semi-finite W$^*$-algebra $\N$ with a faithful normal semi-finite trace $\tau:\N^+\ra[0,\infty]$, let $\MMM(\N,\tau)$ denote the topological $*$-algebra of $\tau$-measurable operators affiliated with $\N$. For any Orlicz function $\orlicz$, a noncommutative Orlicz space $(L_\orlicz(\N,\tau):=\{x\in\MMM(\N,\tau)\mid\exists\lambda>0\,\tau(\orlicz(\lambda\ab{x}))<\infty\},\n{\cdot}_\orlicz:=\inf\{\lambda>0\mid\tau(\orlicz(\lambda^{-1}\ab{x}))\})$ \cite{Kunze:1990}, is uniformly convex and uniformly Fr\'{e}chet differentiable if{}f \cite{Kostecki:2017} (($\N$ is of type II$_\infty$ and $\orlicz,\orlicz^\young\in\mathrm{UC}(\RR)\cap\triangle_2$) or ($\N$ is of type II$_1$ and $\orlicz,\orlicz^\young\in\mathrm{UC}^\infty\cap\triangle_2^\infty\cap\mathrm{SC}(\RR)$) or ($\N$ is of type I and $\Phi\in\triangle_2^0\cap\mathrm{UC}([0,\Phi^{-1}(\frac{1}{2})])$ $\forall\Phi\in\{\orlicz,\orlicz^\young\}$)), where $\orlicz^\young(y):=\sup\{x\ab{y}-\orlicz(x)\mid x\geq0\}$.\footnote{Let $I\subseteq\RR$ be an interval. We call $\orlicz:\RR\ra\RR$ to be \df{Orlicz} if{}f it is convex, with $\orlicz(0)=0$, $\orlicz\not\equiv0$, and $\orlicz(-u)=\orlicz(u)$. An Orlicz $\orlicz$ belongs to: $\triangle_2$ if{}f $\sup_{u>0}\frac{\orlicz(2u)}{\orlicz(u)}<\infty$; $\triangle_2^\infty$ if{}f $\limsup_{u\ra\infty}\frac{\orlicz(2u)}{\orlicz(u)}<\infty$; $\triangle_2^0$ if{}f $\lim_{u\ra0}\frac{\orlicz(2u)}{\orlicz(u)}<\infty$; $\mathrm{UC}(I)$ if{}f $\forall a\in]0,1[$ $\exists\delta(a)\in]0,1[$ $\forall u\in I$ $f(\frac{u+av}{2})\leq\frac{1}{2}(1-\delta(a))(f(u)+f(v))$; $\mathrm{UC}^0$ (resp., $\mathrm{UC}^\infty$) if{}f $\exists u_0>0$ such that $\orlicz\in\mathrm{UC}([0,u_0])$ (resp., $\orlicz\in\mathrm{UC}([u_0,\infty[)$); $\mathrm{SC}(I)$ if{}f $\orlicz$ is strictly convex on $I$.} So, $\Psi_{\varphi,\orlicz}:=\int_0^{\n{\cdot}_\orlicz}\dd t\varphi(t):L_\orlicz(\N,\tau)\ra\RR^+$ is LSQ-adapted and RSQ-compositional. Introducing noncommutative Kaczmarz map $\ell_\orlicz:\N_\star^+\ni\phi\mapsto\orlicz^{-1}(h_\phi)\in L_\orlicz(\N,\tau)^+$, where $\phi=\tau(h_\phi\,\cdot\,)$, we obtain the family $D_{\orlicz,\varphi}:=D_{\ell_\orlicz,\Psi_{\varphi,\orlicz}}$. Due to \cite{Ayupov:Chilin:Abdullaev:2012}, it is independent of $\tau$. For $\N=L_\infty(\X,\mu)$, $\tau=\int_\X\mu$, $\orlicz'(t)>0$ $\forall t>0$ ($\orlicz':=\frac{\dd\Orlicz}{\dd t}$), $\n{\omega}_1=\n{\phi}_1=1$, $\varphi(t)=t^{1/\beta-1}/\beta$, $\beta\in]0,1[$, and $\bar{\orlicz}(\omega,\phi):=\int_\X\mu\orlicz^{-1}(\omega)\orlicz'(\orlicz^{-1}(\phi))$, this gives $D_{\orlicz,\beta}(\omega,\phi)=\frac{1}{\beta}-\frac{1}{\beta}\bar{\orlicz}(\omega,\phi)/\bar{\orlicz}(\varphi,\varphi)$.

\textbf{\hypertarget{Section4Ex4}{Ex.4}.} For a Hilbert space $\H$, $\dim\H=:n<\infty$, Umegaki's information $D_1(\rho,\phi):=\tr_\H(h_\rho(\log h_\rho-\log h_\phi)-h_\rho-h_\phi)$ \cite{Umegaki:1961} equals $D_{\ell=\id,\Psi=\Phi\circ\lewis}(\rho,\phi)$, where $\psi=\tr_\H(h_\psi\cdot)\in\BH_\star^+$, $\lewis$ is a nonincreasing rearrangement of eigenvalues, while $\Phi(x):=\sum_{i=1}^n(x_i\log(x_i)-x_i)$ for $x\geq0$ and $\infty$ otherwise \cite{Bauschke:Borwein:1997}. (This extends to a separable $\dim\H=\infty$ case via \cite{Borwein:Read:Lewis:Zhu:1999}.) So, L\"{u}ders' and quantum Jeffrey's rules \cite{Hellmann:Kaminski:Kostecki:2016}, partial trace \cite{MunkNielsen:2015}, and (preduals of) conditional expectations \cite{Kostecki:Munk:2021}, as special cases of $\RPPP^{D_1}$, belong to $\RPPP^{D_{\ell,\Psi}}$.

\noindent\section{Categories}\hypertarget{Section5}{}

In what follows, $Z=\intefd{\Psi}$. We define the category $\lCvx(\ell,\Psi)$ (resp., $\lAff(\ell,\Psi)$), with objects given by $\ell$-closed $\ell$-convex (resp., $\ell$-closed $\ell$-affine) subsets of $U$, including an $\varnothing$, morphisms given by left $D_{\ell,\Psi}$-projections onto $\ell$-closed $\ell$-convex (resp., $\ell$-closed $\ell$-affine) subsets of these subsets (i.e., $\mathrm{Hom}_{\lCvx(\ell,\Psi)}(\cdot,C)$ consists of $\LPPP^{D_{\ell,\Psi}}_K$ with $K$ varying over all $\ell$-closed $\ell$-convex subsets of $C$), including $\varnothing$ (resulting in empty arrows, e.g., $\varnothing\in\Hom_{\lCvx(\ell,\Psi)}(C_1,C_2)$), and composition given by $\LPPP^{D_{\ell,\Psi}}_{Q_1}\diamond\LPPP^{D_{\ell,\Psi}}_{Q_2}:=\LPPP^{D_{\ell,\Psi}}_{Q_1\cap Q_2}$.\footnote{The composition rule $\diamond$ for left $D_\Psi$-projections is well defined and stable also in the computational sense. Its quantitative evaluation can be performed by means of an algorithm given in \cite{Bauschke:Combettes:2003} (valid for any countable family $\{K_i\}_{i\in I}$ and any $\Psi$ that is totally convex on bounded sets, hence, in particular, for any LSQ-adapted $\Psi$), or by means of  \cite{Bauschke:Lewis:2000,Bregman:Censor:Reich:1999,Dhillon:Tropp:2007} (valid for $\dim X<\infty$, a finite family $\{K_i\}_{i\in\{1,\ldots,n\}}$, and Legendre $\Psi$ satisfying some additional conditions). For $X$ given by the Hilbert space $\H$ and $\Psi_{1/2}=\frac{1}{2}\n{\cdot}_\H^2$, the former algorithm turns to Haugazeau's \cite{Haugazeau:1968} algorithm, while the latter turns to Dykstra's algorithm \cite{Dykstra:1983,Boyle:Dykstra:1986,Han:1988} (valid also for $\dim\H=\infty$, and extendable to countable families $\{K_i\}_{i\in I}$ \cite{Hundal:Deutsch:1997}). Under further restriction of $\{K_i\}$ to a finite family of closed linear subspaces of $\H$, $\LPPP^{D_{\Psi_{1/2}}}_{K_i}$ turn into orthogonal projection operators $P_{K_i}:\H\ra K_i$, while Dykstra's algorithm turns into Halperin's theorem \cite{Halperin:1962} on strong convergence of a cyclic repetition of $P_{K_n}\cdots P_{K_1}$ to $P_{K_1\cap\ldots\cap K_n}$, i.e., $\lim_{k\ra\infty}\n{\left((P_{K_n}\cdots P_{K_1})^k-P_{K_1\cap\ldots\cap K_n}\right)\xi}_\H=0$ $\forall\xi\in\H$. When only two projections are considered, corresponding to a composition $\LPPP^{D_{\Psi_{1/2}}}_{K_1}\diamond\LPPP^{D_{\Psi_{1/2}}}_{K_2}$ for linear subspaces $K_1$ and $K_2$, this becomes the von Neumann--Kakutani theorem \cite{vonNeumann:1933,Kakutani:1940:Nakano}. All these algorithms provide evaluation of the (finite or countable) left $D_\Psi$-projection $\LPPP^{D_\Psi}_{K_1\cap\ldots\cap K_i}(x)$ in terms of a norm convergence of a cyclic sequence of algorithmic steps to the unique limit point. The differences in definitions of those algorithms correspond to different ranges of generality. In particular, while the direct extension on the von Neumann--Kakutani algorithm to closed convex sets converges weakly to an element in the nonempty intersection of $K_1$ and $K_2$ \cite{Bregman:1965} (Kaczmarz's algorithm \cite{Kaczmarz:1937} is a special case of this extension, obtained for hyperplanes and $\dim\H<\infty$), the limit point may be not equal to a projection onto $K_1\cap K_2$ \cite{Combettes:1993} and the norm convergence generally does not hold \cite{Hundal:2004}, although the latter holds always for $\dim\H<\infty$, and can be guaranteed under additional conditions for $\dim\H=\infty$ \cite{Gurin:Polyak:Raik:1967}. On the other hand, the direct extension of Halperin's theorem to linear  projections, of norm equal to 1, onto subspaces of uniformly convex Banach space is norm convergent and returns a projection, of norm equal to 1, onto an intersection \cite{Bruck:Reich:1977}. For noncyclic algorithms, see \cite{Browder:1958,Prager:1960,Amemiya:Ando:1965,Bruck:1982,Hundal:Deutsch:1997,Bauschke:Borwein:1997,Bauschke:Combettes:2003}.} Legendre transform allows us to define the categories $\rCvx(\ell,\Psi)$ (resp., $\rAff(\ell,\Psi)$), with objects given by $(\DG\Psi\circ\ell)$-closed $(\DG\Psi\circ\ell)$-convex (resp., $(\DG\Psi\circ\ell)$-closed $(\DG\Psi\circ\ell)$-affine) subsets of U, including $\varnothing$, morphisms given by right $D_\Psi$-projections onto $(\DG\Psi\circ\ell)$-closed $(\DG\Psi\circ\ell)$-convex (resp., $(\DG\Psi\circ\ell)$-closed $(\DG\Psi\circ\ell)$-affine) subsets of these subsets, including $\varnothing$, and composition given by $\RPPP^{\widetilde{D}_\Psi}_{K_2}\diamond\RPPP^{\widetilde{D}_\Psi}_{K_1}:=\ell^{-1}\circ\DG\Psi^\lfdual\circ\left(\LPPP^{D_{\Psi^\lfdual}}_{(\DG\Psi\circ\ell)(K_2)}\diamond\LPPP^{D_{\Psi^\lfdual}}_{(\DG\Psi\circ\ell)(K_1)}\right)\circ\DG\Psi\circ\ell$. Following Jaynes \cite{Jaynes:1979:where:do:we:stand,Jaynes:2003}, we consider an empty (resp., identity) arrow as an inference corresponding to overdetermination (resp., underdetermination) of constraints. $K_1\cap K_2=K_2\cap K_1$ implies commutativity of $\diamond$. Under restriction of composition by the condition $K_2\subseteq K_1$, the infinitary algorithmic aspect of computation of $\diamond$ can be dropped, defining the convenient categories $\lCvx^\subseteq(\ell,\Psi)$, $\lAff^\subseteq(\ell,\Psi)$, $\rCvx^\subseteq(\ell,\Psi)$, $\rAff^\subseteq(\ell,\Psi)$. On the other hand, by dropping down $\ell$-embeddings everywhere (i.e., moving to $D_\Psi$-projections on $X$), we obtain the categories $\lCvx(\Psi)$, $\lAff(\Psi)$, $\rCvx(\Psi)$, $\rAff(\Psi)$, as well as their $^\subseteq$-subcategories.\footnote{If  $X$ is separable, then $\lAff(\Psi)$ has objects given by the countable sets of polynomial equations as \textit{data types}, and morphisms given by \textit{programs} (algorithms) that translate them (their solutions). More generally, if $X$ is a separable Banach space, then every convex closed subset $C\subseteq X$ is the intersection of the countable number of its supporting closed half-spaces \cite{Bishop:Phelps:1963}, i.e., it is a (countable) polyhedron, which is the set of solutions for a countable system of linear inequalities (see \cite{Borwein:Vanderwerff:2004} for a discussion of the nonseparable case). Hence, also $\lCvx(\Psi)$, at least in the separable case, can be represented as a category of specific data types and computations between them. The resource theory (iii$^\mathrm{L}_{\ell,\Psi}$) from \hyperlink{Section3}{\S3} can be recast as a subcategory $\lCvx_K(\ell,\Psi)$ of $\lCvx(\ell,\Psi)$, determined by the choice of its terminal object to be given by $K$ (so the left $D_{\ell,\Psi}$-projections onto subsets of $K$ are not considered). In such case, the free sets of every object $A$ in $\lCvx_K(\ell,\Psi)$ correspond to the set $\Hom_{\lCvx_K(\ell,\Psi)}(K,A)$, which can be seen as an analogue of the fact that $\Hom_{\texttt{Ban}_\RR}(\RR,Z)$ is equal to the unit ball of a real Banach space $Z$ \cite{Cigler:Losert:Michor:1979}, where $\RR$ is a terminal (and also initial, hence zero) object in the category $\texttt{Ban}_\RR$ of real Banach spaces and completely $\n{\cdot}$-nonexpansive maps \cite{Shvarc:1963}. Each $K\in\Ob(\lAff(\Psi))$ with $\mathrm{codim}(K)=1$ determines a hyperplane in $X$, which can be seen as a resource witness.}

Composability of LSQ$(\Psi)$ (resp., RSQ$(\Psi)$) maps allows to define the category $\LSQ(\Psi)$ (resp., $\RSQ(\Psi)$) of subsets of $\intefd{\Psi}$, with elements of LSQ$(\Psi)$ (resp., RSQ$(\Psi)$) as arrows between them, including empty set as object and empty arrows as morphisms 
(via $\ell$-embedding, this gives $\LSQ(\ell,\Psi)$ (resp., $\RSQ(\ell,\Psi)$)). In general, the composition in $\LSQ(\Psi)$, $\LSQ(\ell,\Psi)$, $\RSQ(\Psi)$ and $\RSQ(\ell,\Psi)$ is not commutative, and their objects are not convex in any sense. Restriction of $\LSQ(\Psi)$ (resp., $\LSQ(\ell,\Psi)$; $\RSQ(\Psi)$; $\RSQ(\ell,\Psi)$) to objects given by the closed convex (resp., $\ell$-closed $\ell$-convex; $\DG\Psi$-closed $\DG\Psi$-convex; $(\DG\Psi\circ\ell)$-closed $(\DG\Psi\circ\ell)$-convex) sets determines a subcategory $\LSQcvx(\Psi)$ (resp., $\LSQcvx(\ell,\Psi)$; $\RSQcvx(\Psi)$; $\RSQcvx(\ell,\Psi)$). Taking subsets of $X$ as objects and elements of CN$(\Psi)$ as arrows defines $\texttt{CN}(\Psi)$ (and $\texttt{CN}(\ell,\Psi)$, via $\ell$). If $X$ is a real Hilbert space and $\Psi=\frac{1}{2}\n{\cdot}_X^2$, then $\texttt{CN}(\Psi)$ coincides with the category $\texttt{Hilb}_{\RR}(X)$ of real Hilbert subspaces of $X$ and completely $\n{\cdot}$-nonexpansive maps. From \hyperlink{Section4Ex2b}{\S4.Ex.2(b)} it follows that $\texttt{CN}(\ell_{1/\gamma},\Psi_{\varphi_\gamma})$, $\gamma\in]0,1[$, coincides with the category of all CPTP maps on $\N_\star$. 

In analogy to Chencov's approach \cite{Chencov:1965,Chencov:1972,Morozova:Chencov:1989} (generalising Blackwell's \cite{Blackwell:1951} statistical equivalence), we will call the subsets $M_1$ and $M_2$ of $\ell^{-1}(\intefd{\Psi})$ to be \df{left equivalent} if{}f $\exists T_1,T_2\in\Arr(\LSQ(\ell,\Psi))$ such that $T_1(M_1)=M_2$ and $T_2(M_2)=M_1$. Hence, the families of left equivalent subsets of $U$ coincide with the groupoids inside $\LSQ(\ell,\Psi)$. Let $\LSQ(\Theta,\ell,\Psi)$ be a subcategory of $\LSQ(\ell,\Psi)$ such that each of its objects is bijectively parametrised by a set $\Theta$. Given $M_1,M_2\subseteq U$, and a set $\Theta$, assume that there exist bijections $\theta_1:\Theta\ra M_1$ and $\theta_2:\Theta\ra M_2$. Adapting linear positive constructions of \cite{LeCam:1964,Raginsky:2011}, we define: a \df{left $(\epsilon,D_{\ell,\Psi})$-deficiency} of $M_2$ with respect to $M_1$ as existence of such $T\in\Hom_{\LSQ(\Theta,\ell,\Psi)}(M_1,\,\cdot\,)$ that $\sup_{\theta\in\Theta}D_{\ell,\Psi}(\theta_2(\theta),T(\theta_1(\theta)))\leq\epsilon$; a \df{left $D_{\ell,\Psi}$-deficiency} of $M_2$ with respect to $M_1$ as $\delta_{D_{\ell,\Psi}}(M_2,M_1):=\inf_{T\in H}\sup_{\theta\in\Theta}D_{\ell,\Psi}(\theta_2(\theta),T(\theta_1(\theta)))$, where $H:=\Hom_{\LSQ(\Theta,\ell,\Psi)}(M_1,\,\cdot\,)$; a \df{mutual left $D_{\ell,\Psi}$-deficiency} of $M_1$ and $M_2$ as $\bar{\delta}_{D_{\ell,\Psi}}(M_1,M_2):=\max\{$ $\delta_{D_{\ell,\Psi}}(M_2,M_1),\delta_{D_{\ell,\Psi}}(M_1,M_2)\}$ (by definition, it is symmetric). Given $M_1,M_2,M_3\in\Ob(\LSQ(\Theta,\ell,\Psi))$, if $\Hom_{\LSQ(\Theta,\ell,\Psi)}(M_1,M_2)\neq\varnothing$ then $\delta_{D_{\ell,\Psi}}(M_3,M_2)\leq\delta_{D_{\ell,\Psi}}(M_3,M_1)$. If $M_1$ and $M_2$ are left equivalent, then $\bar{\delta}_{D_{\ell,\Psi}}(M_1,M_2)=0$. Hence, all objects of a single groupoid in $\LSQ(\Theta,\ell,\Psi)$ have zero mutual left $D_{\ell,\Psi}$-deficiency, yet the latter is nonzero between any elements of two distinct groupoids.\footnote{While the cyclic algorithms mentioned in \hyperlink{Section5}{\S5} exhibit norm convergence in $X$, one still may need either to have a refined quantification of the exactness of intermediate steps, or to quantify the convergence of algorithms with worse convergence behaviours. In such cases left $(\epsilon,D_{\ell,\Psi})$-deficiency can be used to quantify the approximate exactness of a $k$-th cycle of computation of a left $D_{\ell,\Psi}$-projection onto (finite or countable) intersection $M_2=K_1\cap\ldots\cap K_i$, or, more generally, any cyclic convergence algorithm, given $k\in\NN$, with $T:=S^k$, where $S\in\LSQ(\Psi)$ with $S:M_1\ra\intefd{\Psi}$. This illustrates a key property of $D_\Psi$ that underlies the flexibility of its applications: it allows to quantify both algorithmic and structural aspects of the suitable category of spaces, serving as a control interface between arithmetic and geometric layers of a theory.} All of these constructions have their right versions.

The existence and uniqueness of $\LPPP^{D_\Psi}_Q(y)$ does not require norm boundedness (and thus weak compactness) of $Q$, due to coercivity of $D_\Psi(\cdot,y)$ (c.f. Remark 2.13 in 4th ed. of \cite{Barbu:Precupanu:1978} and Lemma 7.3.(v) in \cite{Bauschke:Borwein:Combettes:2001}). Nevertheless, we can consider a subcategory $\lCmpCvx(\Psi)$ of $\lCvx(\Psi)$, consisting of norm bounded, norm closed, convex (equivalently: convex and weakly compact) subsets of $X$ as objects and left $D_\Psi$-projections onto their subobjects as arrows (including empty set and empty arrows). The corresponding $\texttt{r}$-, $\ell$-, and $^\subseteq$- versions of this category are defined analogously as for $\lCvx(\Psi)$. For every $K\in\Ob(\lCmpCvx(\ell,\Psi))$ we can canonically associate an order unit Banach space $A(K)$ of all continuous real valued affine functions on $K$ \cite{Kadison:1951:AMS}, as well as a base norm space $(A(K))^\star$, together with an affine homeomorphism of $K$ onto the base of $(A(K))^\star$ (extending to a linear isomorphism of $\bigcup_{n=1}^\infty n\co(K\cup-K)$ onto $(A(K))^\star$) \cite{Edwards:1964}, as well as a canonical embedding of $A(K)$ into an order unit Banach space $(A(K))^\star{}^\star$ \cite{Ellis:1964} (the latter is equal to the space of all bounded real valued affine functions on $K$ with the supremum norm). Hence, each $K\in\Ob(\lCmpCvx(\ell,\Psi))$ determines a convex operational model in the sense of \cite{Davies:Lewis:1970} (which is a special case \cite{Gudder:1973} of Mielnik's theory of \textit{linear} transmitters \cite{Mielnik:1969,Mielnik:1974}). In consequence, $\lCmpCvx(\ell,\Psi)$ provides a specific nonlinear analogue of the category of convex operational models and positive linear maps with positive duals considered in \cite{Barnum:Duncan:Wilce:2013}.

\noindent\section{Functors}\hypertarget{Section6}{}

We assume $\intefd{\Psi}=X$. From above definitions it follows that every $\ell$ determines a (family of) functor(s), acting by $K\ra\ell(K)$ on objects and $T\mapsto T^\ell$ on arrows, which, together with the functor $\ell^{-1}$, establishes the equivalences of corresponding categories. If $\Psi$ is LSQ-adapted (resp., RSQ-adapted), then an embedding functor $\iota^\mathrm{L}_\Psi:\lCvx(\Psi)\hookrightarrow\LSQcvx(\Psi)$ (resp., $\iota^{\mathrm{R}}_\Psi:\rCvx(\Psi)\hookrightarrow\RSQcvx(\Psi)$) and an induced functor $\iota^\mathrm{L}_{\ell,\Psi}:=\ell^{-1}\circ\iota^\mathrm{L}_\Psi:\lCvx(\ell,\Psi)\hookrightarrow\LSQcvx(\ell,\Psi)$ (resp., $\iota^{\mathrm{R}}_{\ell,\Psi}:=\ell^{-1}\circ\iota^{\mathrm{R}}_\Psi:\rCvx(\ell,\Psi)\hookrightarrow\RSQcvx(\ell,\Psi)$) are well defined, due to $\Fix(\LPPP^{D_\Psi}_{Q_1}\diamond\LPPP^{D_\Psi}_{Q_2})=\Fix(\LPPP^{D_\Psi}_{Q_1})\cap\Fix(\LPPP^{D_\Psi}_{Q_2})=Q_1\cap Q_2$. Given any set $Y$, let $\Pow(Y)$ denote the category of all subsets of $Y$ with functions between them as morphisms. Consider a map $\overline{\co^\mathrm{L}_\Psi(\cdot)}^{w}:\Ob(\Pow(X))\ra\Ob(\Pow(X))$, assigning to each subset $Y$ of a Banach space $X$ the closure of a convex hull $\co(Y)$ of $Y$ in the weak topology of $X$ (it coincides with the norm closure of $\co(Y)$). Let $\overline{\co^\mathrm{L}_\Psi(\cdot)}^{w}:\Arr(\Pow(X))\ra\Arr(\Pow(X))$ be a map that assigns to each function $f:Y_1\ra Y_2$ a map $\LPPP^{D_\Psi}_Q:\overline{\co^\mathrm{L}_\Psi(Y_1)}^{w}\ra\overline{\co_\Psi(Y_2)}^{w}$, where $Q=\overline{\co^\mathrm{L}_\Psi(f(Y_1))}^{w}$. Then $\overline{\co^\mathrm{L}_\Psi(\cdot)}^{w}:\Pow(X)\ra\lCvx(\Psi)$ is a functor. Let $\overline{\co^{\mathrm{R}}_\Psi(\cdot)}^{w}$ be a functor assigning: to each $Y\in\Ob(\Pow(X))$ an image of $\DG\Psi^\lfdual$ of the weak closure of the convex hull of $\DG\Psi(Y)$; to each $f:Y_1\ra Y_2$ a map $\RPPP^{D_\Psi}_Q:\overline{\co^{\mathrm{R}}_\Psi(Y_1)}^{w}\ra\overline{\co^{\mathrm{R}}_\Psi(Y_2)}^{w}$, where $Q=\overline{\co^{\mathrm{R}}_\Psi(f(Y_1))}^{w}$. With a forgetful functor $\mathrm{Frg}_\Set:\lCvx(\Psi)\ra\Pow(X)$ (resp., $\lCvx(\Psi)\ra\Pow(X)$), defined by forgetting convex and topological structure, we obtain an adjunction $\overline{\co^\mathrm{L}_\Psi(\cdot)}^{w}\adj\mathrm{Frg}_\Set$ (resp., $\overline{\co^\mathrm{R}_\Psi(\cdot)}^{w}\adj\mathrm{Frg}_\Set$). If $\Psi$ is LSQ-adapted, then a mapping $\Fix^{\mathrm{L}}_\Psi$, defined by identity on objects of $\LSQcvx(\Psi)$ and assigning $T\mapsto\LPPP^{D_\Psi}_{\Fix(T)}$ to each $T\in\Arr(\LSQcvx(\Psi))$, is a functor $\LSQcvx(\Psi)\ra\lCvx(\Psi)$, satisfying $\iota^{\mathrm{L}}_\Psi\adj\Fix^{\mathrm{L}}_\Psi$. By composition, we obtain $\iota^{\mathrm{L}}_\Psi\circ\overline{\co^\mathrm{L}_\Psi(\cdot)}^{w}\adj\mathrm{Frg}_\Set\circ\Fix^{\mathrm{L}}_\Psi$. By composition with $\ell$, we obtain the functors $\overline{\co^\mathrm{L}_{\ell,\Psi}(\cdot)}^{\ell}:\Pow(U)\ra\lCvx(\ell,\Psi)$, $\iota^{\mathrm{L}}_{\ell,\Psi}:\lCvx(\ell,\Psi)\ra\LSQcvx(\ell,\Psi)$, $\Fix^{\mathrm{L}}_{\ell,\Psi}:\LSQcvx(\ell,\Psi)\ra\lCvx(\ell,\Psi)$, $\mathrm{Frg}_\Set:\lCvx(\ell,\Psi)\ra\Pow(U)$, and the respective adjunctions. If $\Psi$ is RSQ-adapted, then a mapping $\Fix_\Psi^{\mathrm{R}}$, defined by identity on objects of $\RSQcvx(\Psi)$ and assigning $T\mapsto\RPPP^{D_\Psi}_{\Fix(T)}$ to each $T\in\Arr(\RSQcvx(\Psi))$, is a functor $\RSQcvx(\Psi)\ra\rCvx(\Psi)$, satisfying $\iota^{\mathrm{R}}_\Psi\adj\Fix_\Psi^{\mathrm{R}}$. By composition with $\ell$, we obtain the functor $\Fix_{\ell,\Psi}^{\mathrm{R}}$, and the adjunction $\iota^{\mathrm{R}}_{\ell,\Psi}\adj\Fix^{\mathrm{R}}_{\ell,\Psi}$. The endofunctors $\mathrm{Frg}_\Set\circ\overline{\co^\mathrm{L}_{\ell,\Psi}(\cdot)}^{\ell}$ and $\mathrm{Frg}_\Set\circ\overline{\co^\mathrm{R}_{\ell,\Psi}(\cdot)}^{\ell}$ are monads on $\Pow(U)$, while $\Fix^{\mathrm{L}}_{\ell,\Psi}\circ\iota^{\mathrm{L}}_{\ell,\Psi}$ and $\Fix^{\mathrm{R}}_{\ell,\Psi}\circ\iota^{\mathrm{R}}_{\ell,\Psi}$ are monads on $\lCvx(\ell,\Psi)$ and $\rCvx(\ell,\Psi)$, respectively (see \hyperlink{Section8}{\S8} for further discussion). If $\Psi$ is such that both $\LSQ(\Psi)$ and $\RSQ(\Psi)$ are well defined, and assuming additionally \cite{MartinMarquez:Reich:Sabach:2013:BSN} that $\DG\Psi$ and $\DG\Psi^\lfdual$ are (bounded and uniformly continuous) on bounded sets of $\intefd{\Psi}$ and $\intefd{\Psi^\lfdual}$, respectively, the Legendre maps determine an equivalence of categories, given by a pair of functors: $(\cdot)^\Psi:\RSQ(\Psi)\ra\LSQ(\Psi)$ and $(\cdot)^{\Psi^\lfdual}:\LSQ(\Psi)\ra\RSQ(\Psi)$, acting by $C\mapsto\DG\Psi(C)$ and $K\mapsto\DG\Psi^\lfdual(K)$ on objects, and by conjugations $T\mapsto T^\Psi$ and $T\mapsto T^{\Psi^\lfdual}$ on morphisms, respectively. The same definition of $(\cdot)^\Psi$ and $(\cdot)^{\Psi^\lfdual}$, without extra conditions on $\Psi$, gives an equivalence of $\lCvx(\Psi)$ and $\rCvx(\Psi)$.\footnote{This equivalence may seem trivial, as built into the definition of $\rCvx(\Psi)$. Yet, we see it is as a top of an iceberg: there exist right $D_\Psi$-projections which are not Legendre transforms of the left $D_\Psi$-projections \cite{Bauschke:Macklem:Wang:2011}, the equivalence between LSQ$(\Psi)$ and RSQ$(\Psi)$ classes holds only under special conditions \cite{MartinMarquez:Reich:Sabach:2013:BSN}, and there is an important difference between availability of LSQ- vs RSQ-adaptedness in models. Furthermore, while $\LPPP^{D_\Psi}$ correspond to Sanov-type theorems \cite{Sanov:1957,Bjelakovic:Deuschel:Krueger:Seiler:SiegmundSchulze:Szkola:2005,Leonard:2010:entropic}, $\RPPP^{D_\Psi}$ correspond to minimum contrast (e.g., maximum likelihood) estimation \cite{Chencov:1968,Chencov:1972,Eguchi:1983,Amari:Nagaoka:1993}. In general, the dichotomy between $\LPPP^{D_\Psi}$ and $\RPPP^{D_\Psi}$ can be seen as $D_\Psi$-version of a left/right split of a characteristic property $\s{y-P_Cx,x-P_Cx}_\H\leq0$ $\forall(x,y)\in\H\times C$ \cite{Aronszajn:1950} of metric ($=D_{\Psi_{1/2}}$-) projections $P_C$ onto convex closed subsets $C$ in Hilbert space $\H$ under a passage from $\H$ to Banach spaces (left characterising metric projections \cite{Deutsch:1965,Rubinshtein:1965,Lions:1969}, right characterising completely $\n{\cdot}$-nonexpansive sunny retractions \cite{Bruck:1973,Reich:1973}). This leads us to conjecture that the Legendre transform in br\`{e}gmanian setting, under a suitable choice of categories (e.g., left and right $D_\Psi$-Chebysh\"{e}v sets with some additional properties, guaranteeing the composability of respective $D_\Psi$-projections), is an adjunction, with the above equivalence as a special case. Could it be approached via a nucleus of profunctor, as in \cite{Willerton:2015}?}

\noindent\section{Natural transformations and $\mathrm{Hom}$-monoids}\hypertarget{Section7}{}

Let $[0,\infty]$ denote a category consisting of one object $\bullet$, with morphisms given by the elements of the set $\RR^+\cup\{\infty\}$, and their composition defined by addition \cite{Lawvere:1973}. Let $\texttt{2}$ denote the category consisting of two objects, one arrow between them, and the identity arrows on each of the objects. The category $[0,\infty]^\texttt{2}$ has morphisms of $[0,\infty]$ as objects, commutative squares in $[0,\infty]$ as morphisms, and commutative compositions of these squares as compositions. Let $K_1,K_2,K_3,K,L\in\Ob(\lAff^\subseteq_Q(\Psi))$, $K\subseteq K_2$ and $L\subseteq K_3$. For each $\phi\in Q$, left pythagorean equation implies the commutativity of the diagram \eqref{gpt.qpl.diag}. This defines a contravariant functor $D_\Psi(\phi,\cdot):\lAff^\subseteq_Q(\Psi)\ra[0,\infty]^\texttt{2}$, which naturally extends to a functor $D_\Psi(\phi,\cdot):\lAff^\subseteq(\Psi)\downarrow Q\ra[0,\infty]^\texttt{2}$, where $\lAff^\subseteq(\Psi)\downarrow Q$ denotes a slice category of $\lAff^\subseteq(\Psi)$ over $Q$. For any two categories $\texttt{C}$ and $\texttt{D}$, cartesian closedness of the category $\texttt{Cat}$ of all small categories (with natural transformations as morphisms) implies that any functor $\texttt{C}\ra\texttt{D}^\texttt{2}$ corresponds to a natural transformation in $\texttt{D}^\texttt{C}$. Hence, $Q$ parametrises the family of natural transformations $D_\Psi(\phi,\cdot)$ in the category of functors $\lAff^\subseteq(\Psi)\downarrow Q\ra[0,\infty]$. Dependence of $D_\Psi(\phi,\cdot)$ on $Q$ can be factored out by reducing considerations to singletons $Q=\{\phi\}$ (understood as 0-dimensional closed affine spaces). In (some) analogy to \cite{Baez:Fritz:2014,Gagne:Panangaden:2018}. this allows us to state a problem of characterisation of $D_\Psi$ as a natural transformation $D_\Psi(\phi,\cdot)$.
{\begin{equation}
\xymatrix{%
x \ar@{|->}[rr] \ar@{|->}[d]|{\LPPP^{D_\Psi}_K} &&
\left(\;\bullet\;\; \ar[rrrr]_{D_\Psi(\phi,x)}\right. &&&&
\left.\;\;\bullet\;\right) \\
\LPPP^{D_\Psi}_K(x) \ar@{|->}[rr] \ar@{|->}[d]|{\LPPP^{D_\Psi}_L} &&
\left(\;\bullet\;\; \ar[u]^{0}\ar[rrrr]^{D_\Psi(\phi,\LPPP^{D_\Psi}_K(x))}\right. &&&&
\left.\;\;\bullet \ar[u]|{D_\Psi(\LPPP^{D_\Psi}_K(x),x)}\;\right)\\
\LPPP^{D_\Psi}_L\diamond\LPPP^{D_\Psi}_K(x) \ar@{|->}[rr] &&
\left(\;\bullet\;\; \ar[u]^{0}\ar[rrrr]^{D_\Psi(\phi,\LPPP^{D_\Psi}_L\diamond\LPPP^{D_\Psi}_K(x))}\right. &&&&
\left.\;\;\;\;\bullet \ar[u]|{D_\Psi(\LPPP^{D_\Psi}_L\diamond\LPPP^{D_\Psi}_K(x),\LPPP^{D_\Psi}_K(x))}\;\right).\\
}%
\label{gpt.qpl.diag}
\end{equation}}
Given any $Q\in\Ob(\lCvx(\Psi))$, $\Hom_{\lCvx(\Psi)}(\cdot,Q)$ can be equipped with the structure of a commutative partially ordered monoid \cite{Fakhruddin:1986}, with $\LPPP^{D_\Psi}_{Q_1}\diamond\LPPP^{D_\Psi}_{Q_2}:=\LPPP^{D_\Psi}_{Q_1\cap Q_2}$, $\LPPP^{D_\Psi}_{Q_1}\leq\LPPP^{D_\Psi}_{Q_2}:=Q_1\subseteq Q_2$, and a distinguished zero object, given by $\LPPP^{D_\Psi}_Q$. (Examples of computation of $\diamond$ given in \hyperlink{Section5}{\S5} apply here as well.) Hence, each $\Hom_{\lCvx(\ell,\Psi)}(\cdot,Q)$ forms a resource theory in the sense of \cite{Fritz:2017} (which generalises, in particular, the approaches of \cite{Lieb:Yngvason:1999} and \cite{Devetak:Harrow:Winter:2008}). Viewing the order of extended positive reals as a feature distinct from their composition by addition turns $[0,\infty]$ into a commutative partially ordered monoid (with $x+\infty=\infty=\infty+x$ $\forall x\neq\infty$). Thus, each functor $D_{\Psi}(\phi,\cdot)$ can be seen as a morphism $\Hom_{\lAff^\subseteq_Q(\Psi)}(\cdot,Q)\ra[0,\infty]$ inside the category of commutative partially ordered monoids. (By Legendre duality, right pythagorean equation, and $\ell^{-1}$, the above applies also to categories of $\RPPP^{D_\Psi}$, $\LPPP^{D_{\ell,\Psi}}$, and $\RPPP^{D_{\ell,\Psi}}$.)

\noindent\section{Epistemic (co)monads and epistemic resource theories}\hypertarget{Section8}{}

Lawvere \cite{Lawvere:1963} proposed to consider deductive theories of mathematical structures as categories, with their models  given by functors. If $\texttt{C}$ and $\texttt{D}$ are categories, while $F:\texttt{C}\ra\texttt{D}$ and $G:\texttt{D}\ra\texttt{C}$ are functors, such that $F\adj G$, then one can view \cite{Lawvere:1969} (c.f. \cite{Lambek:Scott:1986,Hofmann:1995}): $\texttt{C}$ as a category of (type theoretic) axiomatisations, with objects given by logical formulas and morphisms given by proofs (deductions), $\texttt{D}$ as a category of (geometric) structures modeling these axioms, $F$ as the semantics (meaning) of  $\texttt{C}$ in $\texttt{D}$, and $G$ as the syntax (formalisation) of  $\texttt{D}$ in $\texttt{C}$. Interpreting syntax as a minimal axiomatisation, $F$ can be viewed as the most efficient solution to the problem posed by $G$, while $G$ can be seen as posing the most difficult problem that $F$ solves. On the other hand, Lawvere \cite{Lawvere:1962}, Chencov \cite{Chencov:1965}, and Morse and Sacksteder \cite{Morse:Sacksteder:1966} introduced the category of statistical inferences, with sets of probability densities (probabilistic models) as objects and positive norm-preserving linear maps as arrows. Chencov's approach (viewing the objects as \cytat{figures} \cite{Chencov:1964,Chencov:1968} and their morphisms (statistical decision rules) as \cytat{movements} \cite{Chencov:1965,Chencov:1972}, with statistical equivalence understood as inner groupoids) was focused at relationships between categorical and geometric structures of statistical models and inferences. In his view, the choice of a particular class of morphisms requires justification (he referred to Wald's \cite{Wald:1939,Wald:1950} decision theory), providing a \textit{selection} of the preferred class of maps with respect to a presumed criteria of optimality (given by the Bayes risk). Parallelly, Jaynes \cite{Jaynes:1979:where:do:we:stand,Jaynes:2003} stressed that: 1) probabilities are states of knowledge, which is conditioned upon in the criteria of intersubjective experimental reproducibility (thus, not completely subjective/personal); 2) the mathematical structure of a theory of \textit{inductive} inference should be derived from (determined by) the criteria (requirements) guaranteeing optimality with respect to a particular \textit{logic} of experimental designs/\textit{types} of testable data (c.f. \cite{Tribus:1969}): for each specific method of inductive inference, there are different \textit{experimental designs} that can be \textit{optimally} analysed with it (e.g., $\chi^2$ test makes no sense for a small sample size, the Bayes--Laplace rule is inapplicable to data given by arithmetic means identifiable with average values, etc). 

Our conclusion from these insights, taking into account the large body of evidence on double-sidedness of relationships between `experimental facts' and `intersubjective beliefs' \cite{Duhem:1906,Spengler:1918,Fleck:1935}, is to: 1) consider pairs of: 1a) inductive inference categories, with geometric structures encoding/determining specific prescriptions of optimal/ideal models and inferences, 1b) experimental design categories, seen as logical (type theoretic), and encoding admissible/ideal types of experimental data and their (experimental) transformations; 2) use adjointness, with syntax given by predictive verification (involving frequentist asymptotics and quantitative control of convergence of algorithmic evaluation) and semantics given by model construction (involving infinitary geometric idealisations of finite data sets).

Any category $\texttt{C}$ with object $X\in\Ob(\texttt{C})$ interpreted as a type of knowledge and morphism $f\in\Arr(\texttt{C})$ interpreted as its transformation will be called an \df{epistemic universe}. Consider two epistemic universes: $\ExpDes$ of \df{experimental designs} (with objects given, e.g., by the sets of experimental configuration settings, morphisms given by the sets of parameters of the experimental operations that transform between these settings, and composition of morphisms $h=g\circ f$ representing experimental identification of `performing operation $h$' with `sequential performing of operations $f$ and $g$') and $\IndInf$ of \df{theoretical designs} (with quantified knowledge/information state spaces as objects and inductive inferences/information processings as morphisms). A functor $I:\ExpDes\ra\IndInf$ will be called a \df{model construction} (or \df{interpretation}) while a functor $P:\IndInf\ra\ExpDes$ will be called a \df{predictive verification}. In scientific \textit{inductive} inference, as opposed to mathematical \textit{deductive} inference, the codomain of \textit{semantics} is given by the category of inferences (and thus the syntax is provided by predictive verification), so ``$\forall$ data $\exists$ inference that models it'' (or: ``whatever is measurable, it has to be made thinkable''). On the other hand, the formula ``$\forall$ inferences $\exists$ data that models it'' (or: ``whatever is thinkable, it has to be made measurable'') is characteristic to magical thinking. In consequence, a predictive verification $P$ will be called \df{scientific} (resp., \df{magic}) if{}f $P$ is right (resp., left) adjoint to $I$.\footnote{\cytat{Now, if it comes to making truth, magic can do it far more quickly and brillantly than science. Magic is an experiment in omnipotence; it thinks to create facts by invoking them, as Absolute Will thinks to create truths by assuming them; so after all we need not be surprised that Faust finds magic the best key to the universe} \cite{Santayana:1915}. An adjoint triple $P_m\adj I\adj P_s$ determines a pair $I\circ P_m\adj I\circ P_s$ of monad and comonad on $\IndInf$ and a dual pair $P_m\circ I\adj P_s\circ I$ of comonad and monad on $\ExpDes$, allowing for further interpretation along these lines.} Thus, in \textit{scientific} inductive inference, $\IndInf$ plays a role of a geometric category (answering to a question \cytat{\textit{Whose} information?} \cite{Jaynes:1979:where:do:we:stand}), while $\ExpDes$ plays a role of a type theoretic category (answering to \cytat{Information about \textit{what}?} \cite{Bell:1990}). However: 1) given the fixed choice of two categories, there can be various adjoint pairs of functors between them; 2) the experimental design (`facts') of some family of agents can be a theoretical design (`beliefs') for some other family of agents, and so on. These issues can be (partially) addressed by moving to (co)monads. A choice of a monad (dually: a comonad) on epistemic universe $\texttt{C}$ determines the class of epistemic universes $\texttt{D}$ and corresponding adjoint pairs $I \adj P$ that make $\texttt{C}$ (resp., $\texttt{D}$) to be $\ExpDes$ (resp., $\IndInf$) (dually: $\IndInf$ (resp., $\ExpDes$)). We will call them \df{epistemic (co)monads}.\footnote{So, an epistemic comonad on $\texttt{C}$ limits the possible universes of intersubjective experimental knowledge (together with the corresponding model construction and predictive verification criteria) that are allowed to be built upon $\texttt{C}$ understood as $\IndInf$. Dually, an epistemic monad on $\texttt{C}$ limits the possible theoretical design categories (``optimal models and inferences''), and their relationship with $\texttt{C}$ understood as $\ExpDes$. This leads to the concept of \df{epistemic strategies} for a given epistemic universe $\texttt{C}$, understood as either choosing the specifically crafted monad and comonad (if they are not already given) or utilising the range of available adjunctions equivalent to the given monad and comonad. For example, aiming at maximisation of syntactic power of $\texttt{C}$ as $\ExpDes$, given a fixed monad on it, one would use the largest possible (i.e., the Eilenberg--Moore) category. Dually, aiming at minimisation of semantic power of $\texttt{C}$ as $\IndInf$, given a fixed comonad on it, one would use coKleisli category.} 

Given a choice of a category $\IndInf$ of inductive inferences, an \df{agent} (resp., \df{coagent}) is identified with a monad $J$ (resp., a comonad $E$) on $\IndInf$, encoding the range of available/allowed individual actions/free operations (resp., individually accepted/constructed `facts'). A pair $(E,J)$ of an agent $J$ and coagent $E$ on $\IndInf$ will be called a \df{subject} (or a \df{user}). We define: an \df{epistemic inference theory} as a triple $(\IndInf,E,J)$; a \df{multi-(co)agent epistemic inference theory} as $\mathcal{U}:=(\IndInf,\{E_i\mid i\in \mathcal{I}\},\{J_j\mid j\in\mathcal{J}\})$ (so, $\mathcal{U}$ becomes \df{multi-user} if{}f there is a fixed bijection $\mathcal{I}\iso\mathcal{J}$). Given a choice of a particular (nonunique) adjoint pair $I\adj P$ representing the epistemic comonad $E$, the epistemic monad $J=(J,\mu,\eta)$ can be functorially mapped along $P$, resulting in a monad $\tilde{J}=(\tilde{J},\tilde{\mu},\tilde{\eta})$ over $\ExpDes$, provided there exists a natural transformation $\alpha:\tilde{J}P\Rightarrow PJ$ such that $\alpha\circ\tilde{\eta}P=P\eta$ and $\alpha\circ\tilde{\mu}P=P\mu\circ\alpha J\circ\tilde{J}\alpha$ \cite{Street:1972}.\footnote{A map $(P,\alpha)$ is called a \df{lax morphism} (and: \df{strict} if{}f $\alpha$ is an identity; \df{weak} if{}f $\alpha$ is an isomorphism), while the inversion of direction of $\alpha$ defines a \df{colax morphism} \cite{Street:1972,Leinster:2004}. Lax (resp., colax) morphism induces a functor between corresponding Eilenberg--More (resp., Kleisli) categories, so the choice among them encodes the choice of an epistemic strategy. Dually, given a representation of an epistemic monad, one can subject an epistemic comonad to a (co)lax morphism along this representation, resulting in a ``doubly epistemic'' comonad, encoding (some information about this) in what sense $\IndInf$, now viewed as an experimental design category, was a theoretical design for even more deeper layer of experimental design.}  In this context, a toy model of a ``collective construction of (a system of) scientific facts'' (in the sense of \cite{Spengler:1918,Fleck:1935}) is: given $\mathcal{U}$, the admitted range of possible experimental design categories is limited by the requirement that a single category  $\ExpDes$ has to admit a collection of adjoint pairs $I_i \adj P_i$ $\forall i\in\mathcal{I}$, implementing the whole corresponding family $\{E_i\mid i\in\mathcal{I}\}$ of comonads of $\mathcal{U}$. Given subjects $(E_i,J_i)$ on $\texttt{C}_i$, $i\in\{1,2\}$, and $p,q\in\{\mathrm{lax},\mathrm{colax}\}$, we define a \df{(p,q)-strategy} a pair $((F_E,\alpha_E),(F_J,\alpha_J))$ of $p$ morphism $(F_E,\alpha_E):(\texttt{C}_1,E)\ra(\texttt{C}_2,\tilde{E})$ and $q$ morphism $(F_J,\alpha_J):(\texttt{C}_1,J)\ra(\texttt{C}_2,\tilde{J})$. Intersubjectivity amounts to relating different subjects in a given theory $\mathcal{U}$. Categorifying Chencov's groupoids of statistical equivalence, we define \df{intersubjective commensurability} of (lax,lax)-strategies as an inner groupoid in 2-category $\texttt{InterSubj}_{\mathrm{lax},\mathrm{lax}}$ of subjects of $\mathcal{U}$ as 0-cells, pairs of (lax,lax)-strategies as 1-cells, and pairs of natural transformations $(\kappa_E,\kappa_J):(F_E,F_J)\ra(\bar{F}_E,\bar{F}_J)$, such that $(\bar{\alpha}_E,\bar{\alpha}_J)\circ(E_2,J_2)(\kappa_E,\kappa_J)=(\kappa_E,\kappa_J)(E_1,J_1)\circ(\alpha_E,\alpha_J)$ as 2-cells $((F_E,\alpha_E),(F_J,\alpha_J))\Rightarrow((\bar{G}_E,\bar{\alpha}_E),(\bar{G}_J,\bar{\alpha}_J))$. The corresponding (lax,colax)-, (colax,lax)-, and (colax,colax)- intersubjective categories and their inner commensurabilities (as well as further special cases, given by specialisation of natural transformations $\alpha$ to be weak or strong) are defined analogously.

Every monad $(T,\mu,\eta)$ on a category $\texttt{C}$ gives rise to a monoid $M_T:=(\mathrm{Nat}(\id_\texttt{C},T),\mu(\cdot\circ\cdot),\eta)$. Hence, if $\IndInf$ has a terminal object $\term$, then, given an agent $J$ on $\IndInf$, one can consider the objects of $\IndInf$ as \df{resource spaces}, with: \df{free resources} given by the objects in $\{\sigma_\term(\term)\in\Ob(\IndInf)\mid \sigma\in\mathrm{Nat}(\id_\IndInf,J)\}$, \df{free operations} given by $M_J$, \df{operations} given by all natural transformations from $\id_\IndInf$ to any agent/monad on $\IndInf$, and \df{resource monotones} given by the maps $r:\Ob(\IndInf)\ra[0,\infty]$ such that $r\circ\sigma_A(A)\leq\sigma_A(A)$ $\forall\sigma\in\mathrm{Nat}(\id_\IndInf,J)$ $\forall A\in\Ob(\IndInf)$. Thus, in presence of $\term$ and of at least one nontrivial resource monotone, every (multi-agent) epistemic inference theory becomes a (multi-agent) resource theory. As opposed to set-theoretic case of \hyperlink{Section3}{\S3}, the collection of all operations may be not a monoid itself (lacking a corresponding agent). Hence, although inspired by \cite{delRio:Kraemer:Renner:2015,delRio:Kraemer:2017} and \cite{Chiribella:2018}, the above setting does not reduce to theirs.\footnote{In particular, if $\IndInf$ is a poset $P$, understood as a category (as in \cite{MacLane:1971}), then monads $J$ correspond bijectively to Moore closures on $P$ (c.f., e.g., \cite{Moore:1997}), which are not the same as submonoids of endomorphisms $\mathrm{End}(P)$. Nevertheless, we have a backwards compatibility with the embeddings of \cite{delRio:Kraemer:Renner:2015,delRio:Kraemer:2017}: an interpretation will be called \df{embedding} if{}f it is full and faithful (meaning:  theory should be capable of interpreting consistently all admitted experimental designs, but not necessarily vice versa). An embedding $F:\texttt{C}\ra\texttt{D}$ will be called: \df{extensive} if{}f $F(\texttt{C})$ is a subcategory of $\texttt{D}$; \df{intensive} if{}f there exists a functor $G:\texttt{D}\ra\texttt{C}$ such that $F\adj G$ with the unit of adjunction being a natural isomorphism. Hence, an intensive embedding can be seen as a translation from more coarse-grained/concrete to more refined/abstract description, and defines a comonad $E$ on $\texttt{D}$.} On the other hand, the monoidal category $(\texttt{C}^{\texttt{C}},\circ,\id_{\texttt{C}})$ is not symmetric, so the above setting cannot be recast in terms of \cite{Coecke:Fritz:Spekkens:2016}.


\textbf{Ex.1.} From \hyperlink{Section6}{\S6} we obtain an epistemic inference theory $(\lCvx(\ell,\Psi),\overline{\co^\mathrm{L}_{\ell,\Psi}(\cdot)}^{\ell}\circ\mathrm{Frg}_\Set,\Fix^{\mathrm{L}}_{\ell,\Psi}\circ\iota^{\mathrm{L}}_{\ell,\Psi})$. Each pair $(\ell,\Psi)$ implements a specific convention of intersubjective knowledge construction and its evaluation, that extracts a particular layer of data from the subsets of $U$, and enriches it with a particular idealisation, corresponding to the chosen quantitative criteria of optimal inference.

\textbf{\hypertarget{Section8Ex2}{Ex.2}.} $(\Pow(\N_\star),\id,\{\mathrm{Frg}_\Set\circ\overline{\co^\mathrm{L}_{\ell_\orlicz,\Psi_\varphi}(\cdot)}^{\ell}\})$, with $\Orlicz$ and $\varphi$ varying as in \hyperlink{Section4Ex3}{\S4.Ex.3}, is a multi-agent epistemic inference theory. Kaczmarz map $L_{\orlicz_1}(\N,\tau)\ni x=u_x\ab{x}\mapsto u_x\orlicz_2^{-1}(\orlicz_1(\ab{x}))\in L_{\Orlicz_2}(\N,\tau)$ is a homemorphism \cite{Kostecki:2017}, seting up categorical equivalences between $\lCvx(\ell_{\orlicz},\Psi_\varphi)$ for varying $\orlicz$ and fixed $\varphi$, implying strict intersubjective commensurability of corresponding monads/agents on $\Pow(\N_\star)$. Each agent corresponds to a family of resource theories of states of type (iii$^{\mathrm{L}}_{\ell_\orlicz,\Psi_\varphi}$), parametrised by $\ell_\orlicz$-closed $\ell_\orlicz$-convex sets of free states.\footnote{Analogous statements hold for corresponding monads/agents on $\Pow(A_\star)$ (resp., $\Pow(\N_\star)$), constructed according to \hyperlink{Section4Ex2a}{\S4.Ex.2(a)} (resp., \hyperlink{Section4Ex2a}{\S4.Ex.2(b)}), via nonassociative (resp., noncommutative) Mazur map $L_p(A,\tau)\ni x\mapsto\sgn(x)\ab{x}^{p/q}\in L_q(A,\tau)$ (resp., $L_p(\N)\ni x=u_x\ab{x}\mapsto u_x\ab{x}^{p/q}\in L_q(\N)$) as homeomorphism \cite{Kostecki:2017} (resp., \cite{Kosaki:1984:uniform,Raynaud:2002}).} On the other hand, $\Pow(\N_\star)$ has a terminal object, allowing to ask: what are the nontrivial resource monotones turning this example in a multi-agent resource theory?

\vspace{20mm}\noindent{}\textbf{Acknowledgments}

\vspace{2mm}
\begin{spacing}{1}\noindent{}{\normalsize I thank L\'{\i}dia del Rio, Tobias Fritz, Karol Horodecki, and Anna Jen\v{c}ov\'{a} for discussions. This research was supported by 2015/18/E/ST2/00327 grant of National Science Centre, as well as by Perimeter Institute for Theoretical Physics. Research at Perimeter Institute is supported by the Government of Canada through Industry Canada and by the Province of Ontario through the Ministry of Research and Innovation. Part of this research was conducted during my visit at Department of Mathematical Informatics, Graduate School of Information Science, Nagoya University, on academic leave from University of Gda\'{n}sk. I wish to express my sincere gratitude to Francesco Buscemi for an invitation, discussions, and kind hospitality.}\end{spacing}

\newpage\section*{References}

\begin{spacing}{0.9}{\small (The following bijective Latin transliteration of Russian Cyrillic script is used: {\cyrrm{ts}} = c, {\cyrrm{ch}} = ch, {\cyrrm{zh}} = zh, {\cyrrm{sh}} = sh, {\cyrrm{shch}} = \v{s}, {\cyrrm{y}} = y, {\cyrrm{i}} = i, {\cyrrm{yu}} = yu, {\cyrrm{ya}} = ya, {\cyrrm{\"{e}}} = \"{e}, {\cyrrm{\cdprime}} = `, {\cyrrm{\cprime}} = ', {\cyrrm{\`{e}}} = \`{e}, {\cyrrm{\u{i}}} = \u{\i}, {\cyrrm{kh}} = kh, and analogously for capitalised letters, with an exception of {\cyrrm{Kh}} = H at the beginnings of words.)}\end{spacing}

\bibliographystyle{../../../rpkbib}
\renewcommand\refname{\vskip -1.8cm}
\bibliography{../../../rpk}

\end{document}